\documentclass[11pt,twoside]{article}

\usepackage{asp2006}
\usepackage{epsf}
\usepackage{psfig}
\usepackage{lscape}

\usepackage{graphicx}

\begin{document}

\title{The Influence of Helical Magnetic Fields in the Dynamics and Emission of Relativistic Jets}
\author{Roca-Sogorb, M.$^{1}$, Perucho, M.$^{2}$, G\'omez, J. L.$^{1}$, Mart\'i, J. M.$^{3}$, \\Ant\'on, L.$^{3}$, Aloy, M. A.$^{3}$ \& Agudo, I.$^{1}$}   
\affil{$^{1}$ Instituto de Astrof\'isica de Andaluc\'ia (CSIC). Granada, Spain.\\
$^{2}$ Max-Planck Institut f\"{u}r Radioastronomie. Bonn, Germany.\\
$^{3}$ Departament d'Astronomia i Astrof\'isica. Universitat de Val$\grave{e}$ncia. Valencia, Spain}

\begin{abstract} We present numerical relativistic magnetohydrodynamic and emission simulations aimed to study the role played by the magnetic field in the dynamics and emission of relativistic jets in Active Galactic Nuclei. We focus our analysis on the study of the emission from recollimation shocks since they may provide an interpretation for the stationary components seen at parsec-scales in multiple sources. We show that the relative brightness of the knots associated with the recollimation shocks decreases with increasing jet magnetization, suggesting that jets presenting stationary components may have a relatively weak magnetization, with magnetic fields of the order of equipartition or below. 
\end{abstract}

\section{Introduction}   

Thanks to the recent developments in the numerical modelling of magnetized relativistic jets it is now possible to study the influence of the magnetic field in the dynamics of relativistic jets. Computation of non-thermal (synchrotron) emission using relativistic magnetohydrodynamic (RMHD) simulations as inputs allow us to obtain synthetic radio maps that can be directly compared with actual observations (see, e.g., G\'omez 2002, and references therein). This comparison between simulations and observations is proving a powerful tool in the understanding of the physical processes taking place in the jets of Active Galactic Nuclei (AGN).

\section{eRMHD Simulations}

The simulations have been performed using a numerical code that solves the RMHD equations in conservative form and cylindrical coordinates with axial symmetry (see Leismann et al. 2005 for more details). Numerical fluxes between contiguous zones are computed using an approximate Riemann solver. Spatial second order accuracy is achieved by means of piecewise linear, monotonic reconstruction of the fluid variables. Averaged values of the conserved variables are advanced in time by means of a Runge-Kutta algorithm of third order. The magnetic field configuration is kept divergence-free thanks to the implementation of a constrained transport method.
Using the RMHD results as inputs we compute synthetic radio continuum emission (total and linearly polarized) maps that can be directly compared with observations. For this we integrate the transfer equation for the synchrotron radiation (for an optically thin frequency and in arbitrary units) taking into account the appropriate relativistic effects. See G\'omez (2002), and references therein, for a complete description.

\begin{figure}[!ht]
\begin{center}
\includegraphics[width=0.73\textwidth]{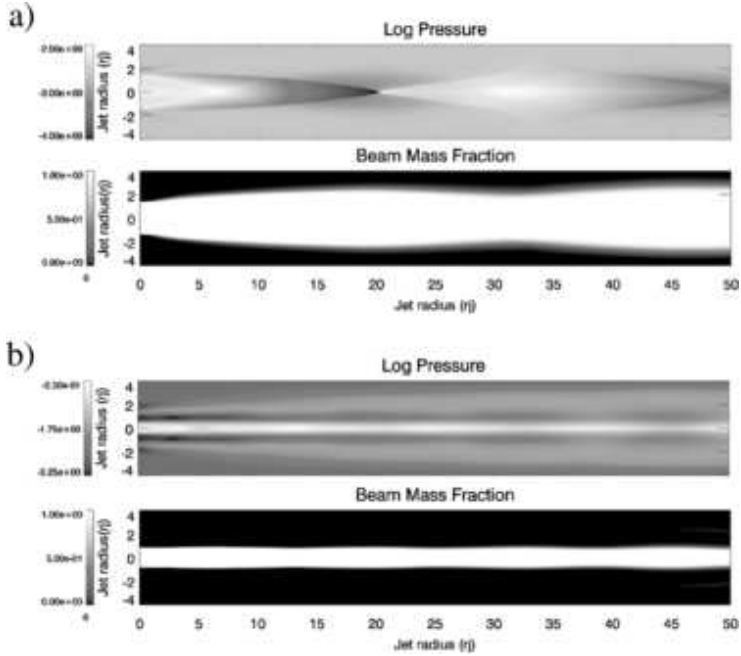}
\caption{Panels show the logarithm of the pressure and the beam mass fraction (ratio of jet to external medium rest-mass densities) for Model A \emph{(a; top)} and Model B \emph{(b; bottom)}.}
\end{center}
\end{figure}

\section{Setting up the numerical simulations}

The models are characterized by the injection velocity into the external medium ( i.e, Lorentz factor $\Gamma$), and the specific internal energy. Two more parameters, the beam to external medium ratios of pressure, $K$, and rest-mass density at the injection position, are used to fix the properties of the atmosphere through which the jet propagates. An ideal gas equation of state is used. We assume that the magnetic field has a helical structure (see Lind et al 1989; Komissarov 1999), with axial, $B^z$, and toroidal, $B^{\phi}$, components given by:
\begin{equation}
B^z = \left\{ 
\begin{array}{rcl}
 B^z & \mbox{if} & r \leq r_j\\
 0     & \mbox{if} & r > r_j
\end{array}
\right\}
\quad,\quad
B^{\phi} = \left\{
\begin{array}{rcl}
 B^{\phi}_m (r/r_m) & \mbox{if} & r  \leq r_m\\
 B^{\phi}_m (r_m/r) & \mbox{if} & r_m \leq r \leq r_j\\
 0                  & \mbox{if} & r > r_j
\end{array}
\right\}
\end{equation}

\noindent where $r_{j}$ is the jet radius, and $r_m$ is the radius at which the toroidal magnetic field reaches its maximum value $B_m^{\phi}$. The magnetic field is characterized by the averaged pitch angle $\phi$, which is defined as $\tan\phi=B^{\phi}/B^z$, and the magnetization parameter $\beta$, which is the ratio of magnetic field to gas pressure.

\section{The Influence of the Jet Magnetization}

In order to study the influence of the magnetization in the dynamics and emission we have performed simulations of overpressured ($K$=2) relativistic ($\Gamma=4$) jets carrying a helical magnetic field with pitch angle $\phi=65\deg$ and two different magnetizations at the jet inlet: Model A with $\beta$=0.1, and Model B with $\beta$=3. 

\begin{figure}[h]
\begin{center}
\includegraphics[width=0.95\textwidth]{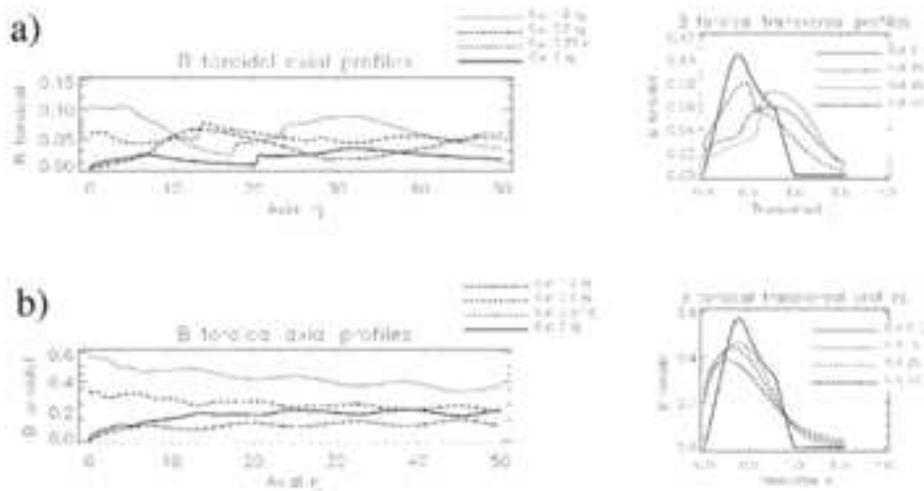}
\caption{Toroidal magnetic field profiles at different radius along and across the jet for Model A \emph{(a; top)} and Model B \emph{(b; bottom)}.}
\end{center}
\end{figure}

The dynamics of the jet is dominated by the tendency of the beam fluid to achieve pressure equilibrium with the external medium, causing overexpansions and overcontractions past equilibrium of the fluid leading to the formation of a pattern of recollimation shocks (see Fig.~1). We can study the role played by the magnetic field in the dynamics by analyzing the influence of the toroidal and axial components separately. The toroidal component (see Fig.~2) is relatively weak close to the jet axis but it can be significant close to its maximum strength. This component generates a force against the expansions and contractions of the fluid, which will be maximum at radius $r_m$. Studying the position of $r_m$ in the transversal profiles at different radius along the jet (see Fig.~2) we see that it changes during the expansions and contractions of the jet, but remains close to the original position for Model A while for Model B moves to an inner radius with distance along the jet. The contribution of the axial component, which has a constant profile inside the jet, is to generate a magnetic tension that opposes to the transversal movement of the jet. Therefore both components contribute to the confinement, collimation, and stability of the jet. Hence a more magnetized jet will not expand initially as much as a jet with a lower magnetization, and the subsequent contraction will take place before in the grid length, resulting in weaker shocks. Certainly, by comparing our models A and B (see Figs.~1 and 2) we observe that i) the recollimation shocks are weaker, and there is larger number of them, as $\beta$ increases; ii) Model B presents a stronger collimation, induced by the enhanced magnetic field.

Figure 3 shows the computed synthetic emission maps for a viewing angle of $14\deg$ for models A and B. We observe bright knots of emission associated with the recollimation shocks, which may provide an interpretation for the stationary components commonly seen in multiple sources (e.g., G\'omez et al. 1995; Agudo et al. 2001; Jorstad et al. 2005). Figure 3 shows that the relative brightness of the knots associated with the internal shocks decreases with increasing magnetization. Indeed, we observe that Model A presents strong stationary components with a relative brightness that doubles the emission of the underlying flow. On the other hand, Model B, which has very weak recollimation shocks (see Fig.~1), shows a smooth jet emission with small variations along the jet. We can therefore suggest that jets presenting stationary components may have a relatively weak magnetization, with $\beta$ close or below equipartition.

\begin{figure}[!ht]
\begin{center}
\includegraphics[width=1.0\textwidth]{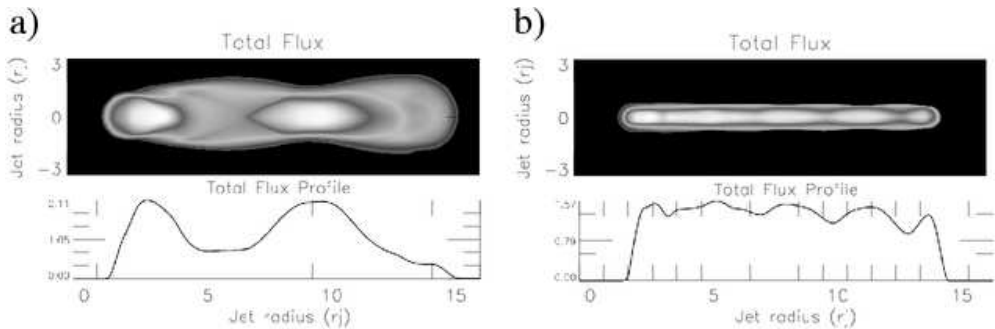}
\end{center}
\caption{Panels show total emission maps in arbitrary units (top) and total emission along the jet axis direction (bottom) for a) Model A and b) Model B for a viewing angle of $14\deg$.}
\end{figure}



\acknowledgements 
This research has been partially supported by the Spanish Ministerio de Educaci\'on y Ciencia and the European Fund for Regional Development through grants AYA2004-08067-C03-03 and AYA2004-08067-C03-01. M.P. acknowledges a 
postdoctoral fellowship of the Generalitat Valenciana (Beca 
Postdoctoral d'Excel$\cdot$l\`encia). I.~A. has been supported by an I3P contract by the Spanish ``Consejo Superior de Investigaciones Cient\'{i}ficas".



\end{document}